\documentclass[%
 aip,
 amsmath,amssymb,
preprint,%
]{revtex4-1}

\usepackage{graphicx}
\usepackage{dcolumn}
\usepackage{bm}
\usepackage{color}
\usepackage{comment}

\usepackage[utf8]{inputenc}
\usepackage[T1]{fontenc}
\usepackage{mathptmx}
\usepackage{etoolbox}
\usepackage{float}

\makeatletter
\def\@email#1#2{%
 \endgroup
 \patchcmd{\titleblock@produce}
  {\frontmatter@RRAPformat}
  {\frontmatter@RRAPformat{\produce@RRAP{*#1\href{mailto:#2}{#2}}}\frontmatter@RRAPformat}
  {}{}
}%
\makeatother
\begin{document}

\preprint{Preprint}

\title{A Composite Hydrogel of Porous Gold Nanorods and Gelatin: Nanoscale Structure and Rheo-Mechanical Properties}

\author{Irfan Khan}
\affiliation{ 
Department of Physics, Indian Institute of Technology Bombay
}
\author{Snigdharani Panda}
\affiliation{ 
Department of Physics, Indian Institute of Technology Bombay
}
\author{Sugam Kumar}
\affiliation{ 
Solid State Physics Division, Bhabha Atomic Research Centre
}
\author{Sunita Srivastava*}
 \email[Author to whom correspondence should be addressed:]{sunita.srivastava@iitb.ac.in}
\affiliation{ 
Department of Physics, Indian Institute of Technology Bombay
}

\date{\today}

\begin{abstract}
Incorporating nanomaterials into hydrogels allows for the creation of versatile materials with properties that can be precisely tailored by manipulating their nanoscale structures, leading to a wide range of bulk properties. Investigating the structural and property characteristics of composite hydrogels is crucial in tailoring their performance for specific applications. This study focuses on investigating the correlation between the structural arrangement and properties of a composite hydrogel of thermoresponsive polymer, gelatin, and light-responsive antimicrobial porous gold nanorods,\textit{PAuNR}.
The rheo-mechanical properties of the composite hydrogels are correlated with their nanoscale structural characteristics, investigated using small-angle neutron scattering (\textit{SANS}). Analysis of \textit{SANS} data reveals a decrease in the fractal dimension of \textit{PAuNRs} incorporated hydrogel matrix, as compared to pure gelatin. Incorporating \textit{PAuNRs} results in formation of softer composite hydrogel as evident from decrease in viscoelastic moduli, critical yield strain, denaturation temperature and swelling ratio. Our results demonstrates that the structural modulation at the nanoscale can be precisely controlled through adjusting  \textit{PAuNRs} concentration and temperature providing an fabrication mechanism for hydrogels with desired elastic properties. The reduced elasticity of the composite hydrogel and light sensitive/antimicrobial property of the  \textit{PAuNRs} makes this system suitable for specific biomedical applications, such as tissue engineering, device fabrication and stimuli based controlled drug delivery devices respectively. 
\end{abstract}

\maketitle

\section{\label{sec:level1}INTRODUCTION}

Hydrogels, network of hydrophillic polymer chains cross-linked together, with their exceptional absorbency and flexibility, are particularly valuable in biomedical applications, printing, and wearable devices  \cite{garcia2015biocompatible,gubaidullin2022modulation,barrett2021multifunctional}. Gelatin, a natural polymer obtained from denatured collagen \cite{wang2021physically}, has become a prominent choice in biomedical applications due to its  excellent biocompatibility and biodegradability properties \cite{liu2020tunable,C3SM00125C,berts2013structure,panda2024nir}. Gelatin-based hydrogel exhibits mechanical and biochemical properties similar to the extracellular matrix (ECM) of the body, making it the most popular choice for application in tissue engineering, cell differentiation, bio-printing, and so on \cite{ma2021mammalian,wang2022three}. The interesting thermo-responsive properties of gelatin arise due to a combination of interatomic interactions, including hydrogen bonding (H-bonding), hydrophobic interactions, and van der Waals forces \cite{van2000structural,wisotzki2017influence}. This unique property makes them a versatile material in various biomedical and environmental applications. The physical properties of the gelatin hydrogel can be modified by incorporation of polymers, cross-linkers, nanoparticles, etc. By modifying hydrogels through cross-linking processes or by incorporating specific entities such as ions or nanoparticles into the hydrogel matrix, the structure and properties of the hydrogel can be precisely tailored to meet the requirements of specific applications \cite{ma2021mammalian,wang2022three}.\par
Incorporating different ions and nanoparticles into gelatin hydrogels is a widely used strategy to tailor their properties  and have shown significant development \cite{qiao2021influence,wu2018ionic}. The trend in hydrogel research is moving away from ion-based materials and towards nanoparticle-based ones. This shift is motivated by the enhanced capabilities that nanoparticles provide, including better control of drug release, stronger mechanical properties, and the ability to perform multiple functions. Nanocomposites allow for more precise, responsive, and versatile applications, making them better suited for advanced biomedical and technological uses. It has been shown that the young modulus of gelatin hydrogel is enhanced by \textit{2.9} times after photocrosslinking of CTAB coated AuNRs (1mg/ml) with gelatin methacrylate. Electrostatic interaction between CTAB-AuNR and gelatin methacrylate responsible not only for higher gel strength but also smaller pore size of the composite hydrogel\cite{navaei2016gold}. Synergistic chemical and physical interactions between porous silica nanoparticles and a matrix of  gelatin/aldehyde-modified xanthan lead to a tougher hydrogel structure \cite{aghajanzadeh2024augmented}. Gelatin hydrogel matrix embedded with 30 wt $\%$ of spindle-like hydroxyapatite (HA)  exhibits a compressive modulus 30.6 lower than pure gelatin hydrogel. These electrostatic interactions between calcium and carboxylate groups limit the sites for cross-linking between polymer chains within the hydrogel network resulting in reduction of gel strength \cite{raucci2018gelatin}. There are several parameters that effect the macroscopic response of the nanoparticle composite gels. To understand how these gels work at a larger scale, we need to know what happens at a smaller, atomic level. Specifically, understanding the internal structure of particle-embedded gelatin hydrogels using  techniques sensitive to nanoscale structure as such small angle neutron scattering  which can reveal the internal structure of these gels at a very small scale and correlating the structure to bulk properties is desirable.\\
Porous nanoparticle-based hydrogel composites have gained significant attention due to their ability to offer controlled and stimuli-responsive applications \cite{Rocha2016}. These materials combine the unique properties of hydrogels (absorbency, biocompatibility) with the enhanced functionalities of nanoparticles such as antibacterial properties, electrical conductivity, or magnetic responsiveness \cite{li2022recent,myrovali2022hybrid}. In this study, we incorporate the freshly synthesized \textit{PAuNRs} samples into gelation hydrogel matrix for formation of gelation-\textit{PAuNRs} hydrogel composite.  Gelatin composite hydrogel is synthesized by
incorporation of Cysteamine-coated porous gold nanorods (\textit{PAuNRs}) along with a fixed amount of tannic acid as cross-linkers. Tannic acid as a physical cross-linker is employed to enhance the mechanical, thermal, and antimicrobial
properties of gelatin hydrogel \cite{leite2021effect,panda2024nir}. Cysteamine-coated \textit{PAuNRs} 
were chosen due to their inherent properties, including porosity, near-infrared (NIR) absorption, conductivity, and
antibacterial activity. These properties endow the gelatin composite hydrogel with functionalities suitable for 
diverse applications, including drug delivery, photoacoustic imaging, computed tomography imaging, and even 
photothermal therapy \cite{cai2018design}. This research investigated how the incorporation of \textit{PAuNRs} with 
different concentrations and variations in temperature influence the architecture of gelatin composite hydrogels.
We then explored the connection between these structural changes with its rheo-mechanical and thermal properties of
the hydrogels. Small-angle neutron scattering (\textit{SANS}) was performed on both pure gelatin and composite hydrogels at
different temperatures to understand the resulting structural modulations. Additionally, rheo-mechanical testing, temperature sweep tests, and DSC measurements were employed to analyze the rheo-mechanical and thermal 
properties of the hydrogels. Understanding the effect of  \textit{PAuNRs} concentration and temperature on structural modulation of hydrogels, and how these structural changes impact macroscopic properties, is crucial for designing hydrogels with suitable characteristics for specific applications.

\section{EXPERIMENTS}

\subsection{Materials}
The various chemicals utilized in this study such as Gelatin type A from porcine skin, selenious acid $(Se{O_2})$, tellurium dioxide $(Te{O_2})$, hydrazine monohydrate, tetrachloroauric $(III)$ acid hydrate $(HAu{Cl_4})$, 2-aminoethane thiol were all purchased from Sigma Aldrich. HPLC grade tannic acid from sigma aldrich is used in for synthesis. Sodium dodecyl sulfate $(SDS)$ was supplied by Emplura. Ultrapure deionized water (18.2 M$\Omega$-cm) was used throughout the experiments. All the chemicals were used as received.

\subsection{Synthesis of Porous Gold Nanorods (\textit{PAuNRs})}

\textit{PAuNRs} synthesis is a two-step process\cite{cai2018design}. The first step involves the reduction of tellurium (Te) and selenium (Se) with hydrazine monohydrate to produce Te nanorods. The synthesis of Te nanorods involved adding tellurium dioxide,  seleniniouc acid, and hydrazine monohydrate to a flask and stirring continuously at 40 $^\circ\mathrm{C}$ for 10 minutes. To stop the reaction and stabilize the nanorods, the solution was then diluted 10 times with SDS. After another 10 minutes of stirring, the mixture was centrifuged three times to isolate the Te nanorods. These Te nanorods served as templates for the subsequent synthesis of \textit{PAuNRs}. In the second step, the Te nanorods are used as templates for the synthesis of \textit{PAuNRs} through a sacrificial galvanic replacement reaction. The Te nanorods were added to an ice-cold solution of cysteamine under constant stirring. After 10 minutes, the mixture was purified by centrifugation, repeated three times.

\subsection{Gelatin nanocomposite hydrogel preparation}
A gelatin hydrogel can be prepared by dissolving 0.2 grams each of gelatin (type A) and tannic acid in 4 milliliters of ultrapure deionized water. This ensures a final concentration of 5\% wt/wt for both components within the gel. The mixture is then stirred  for one hour on magnetic stirrer at constant 550 rpm and 45 $^\circ\mathrm{C}$. After thorough mixing, the solution is carefully poured into 20 mm diameter silicone septa. These septa are then placed in a vacuum desiccator and stored at 4 $^\circ\mathrm{C}$ for 24 hours.
Gelatin nanocomposite (GNC) hydrogel samples were prepared by mixing \textit{PAuNRs} suspensions of varying concentrations (0 mg/ml,
 0.5 mg/ml, 1.25 mg/ml, 2.5 mg/ml, and 5 mg/ml) with a 5\% wt/wt gelatin solution, followed by incubation at room temperature for \textit{24 hours} to allow gelation. These samples were labeled as Gelatin, GNC1, GNC2, GNC3, and GNC4, respectively. For \textit{SANS} measurements,
 the hydrogels were prepared using deuterated water to enhance the contrast between the hydrogel and the surrounding solvent.

\begin{figure}
\centering
\includegraphics[width={\textwidth}]{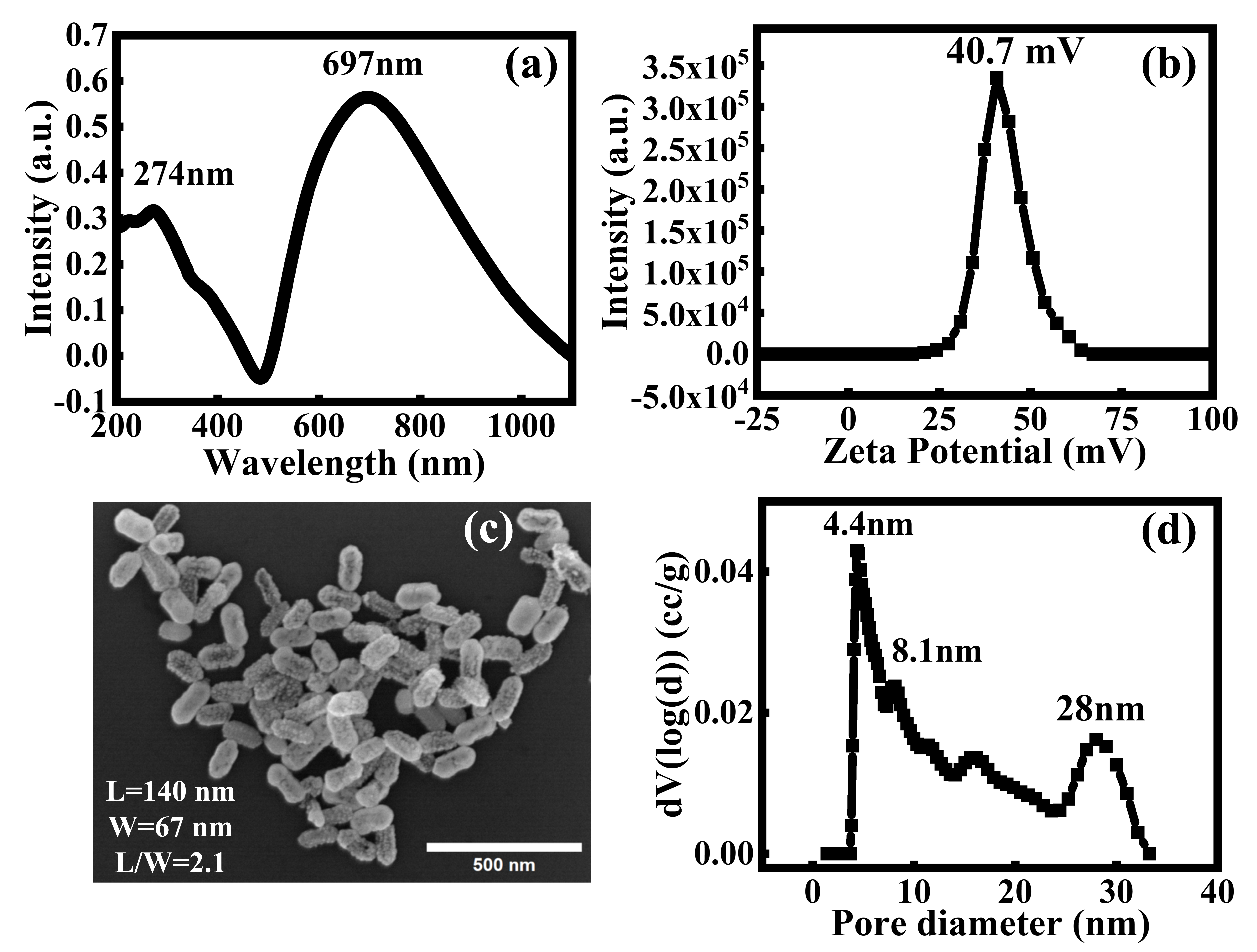}
\caption{(a) UV-Vis absorption spectra of \textit{PAuNR} aqueous suspension, showing two surface plasmon peaks corresponding to longitudinal (697 nm) and transverse (274 nm) length scales. (b) Zeta potential estimate of $+40.7~mV$ indicates an positively charged surface for cystemaine coated \textit{PAuNR} suspensions. (c) SEM images for \textit{PAuNRs} reveal cylindrical-shaped \textit{PAuNR} with a mean length of \textit{140 nm} and width of \textit{67 nm} (inset-size histograms obtained from Imagej analysis). (d) BET analysis pore size distribution plot, demonstrating pore sizes ranging from \textit{4.4 nm} to \textit{28 nm}, with the highest contribution from pores of size \textit{4.4 nm}.} 
\label{fig1}
\end{figure}

\section{RESULTS AND DISCUSSION}
\subsection{Structural Analysis of \textit{PAuNRs} and Gelatine hydrogel-\textit{PAuNRs} composites}

The UV-Vis (ultraviolet-visible) absorption spectra of freshly synthesized \textit{PAuNRs} particles dispersed in aqueous solutions were recorded using a Jasco V-730 double-beam UV-Vis spectrophotometer. Measurements were conducted in the wavelength range of 200 nm to 1100 nm using a 1 cm path length cuvette. A prominent surface plasmon resonance (SPR) peak was observed at \textit{$\sim$ 697 nm} as shown in the spectrum in Fig. 1(a). This peak falls within the first near-infrared (NIR) region, suggesting that these gold nanorods (\textit{PAuNRs}) could be suitable for NIR-triggered drug delivery applications. Further, the characterization of the \textit{PAuNRs} using Zeta potential measurements (Malvern Zetasizer ZS DLS instrument) reveal an surface potential estimate of \textit{+40.7 mV}(Fig. 1(b), which signifies stable dispersion of \textit{PAuNRs} coated with the positively charged 2-aminoethane thiol. The size and aspect ratio estimate of the \textit{PAuNRs} are determined using SEM as shown in Fig. 1(c). The SEM image analysis reveals the formation of a fairly monodisperse rod-shaped structure with an aspect ratio of \textit{$\sim$2.1}. The size estimate by counting over 300 particles were found to be $140 \pm 11$ nm in length and $67 \pm 6$ nm in width (Fig. S1). SEM images revealed a rough surface of the \textit{PAuNRs}, indicating a porous structure, which was confirmed by BET analysis (Fig. 1d). The $N_2$ adsorption-desorption isotherms were recorded on the BET surface area analyzer (Quantachrome) at 77 K. The samples were degassed at 120 $^\circ\mathrm{C}$ for 10 hours under high vacuum conditions. The DFT method\cite{bardestani2019experimental} was applied to measure the pore size distribution from the nitrogen adsorption data. In Fig. 1d we show the pore size distribution by BET analysis, indicating that the pores sizes ranges from 4.40 nm to 28 nm, having a prominent size of 4.40 nm on the \textit{PAuNRs} surface. The \textit{PAuNRs} synthesized in this work is NIR active and the porous structure on the particle surface has application in drug delivery as demonstrated earlier \cite{panda2024nir}. Given the potential applications of the gelation-\textit{PAuNR} composite in controlled drug delivery, light-triggered drug release, photo-thermal imaging, and photo-thermal therapy \cite{cai2018design}, we focus on investigating how \textit{PAuNR} affects the structural, thermal, and mechanical properties of gelatin hydrogels. Understanding the relationship between a material's structure and its properties is essential for tailoring those properties to meet the specific requirements of a particular application. 
\par

The structural morphology of the gelatin-\textit{PAuNR} hydrogel dispersion  was investigated using small-angle neutron scattering (\textit{SANS}) diffractometer at Guide tube laboratory, Dhruva reactor Bhabha Atomic Research Centre (BARC), Mumbai India. A neutron beam of wavelength 5.2 \AA\ (wavelength spread, ${\Delta\lambda/\lambda=15\%}$) was incident on the sample and the scattered neutron beam was detected by 1 m long $1D~ He^{3}$  position sensitive detector. The wave vector, \textit{q} ($(4\pi/\lambda)\sin{\theta/2}$ where $\theta$ is scattering angle)
 range of 0.0157 $\text{\r{A}}^{-1}$ to 0.25 $\text{\r{A}}^{-1}$  were set for all experiments at varying temperature and composition of \textit{PAuNR}. To improve data quality, corrections for background noise, instrument transmission, and contributions from empty cells were applied. The hydrogel samples are prepared using heavy water to minimize incoherent scattering and increase contrast for hydrogenous components such as gelatin polymers. \par

\begin{figure}[h!]
\centering
\includegraphics[width={\textwidth}]{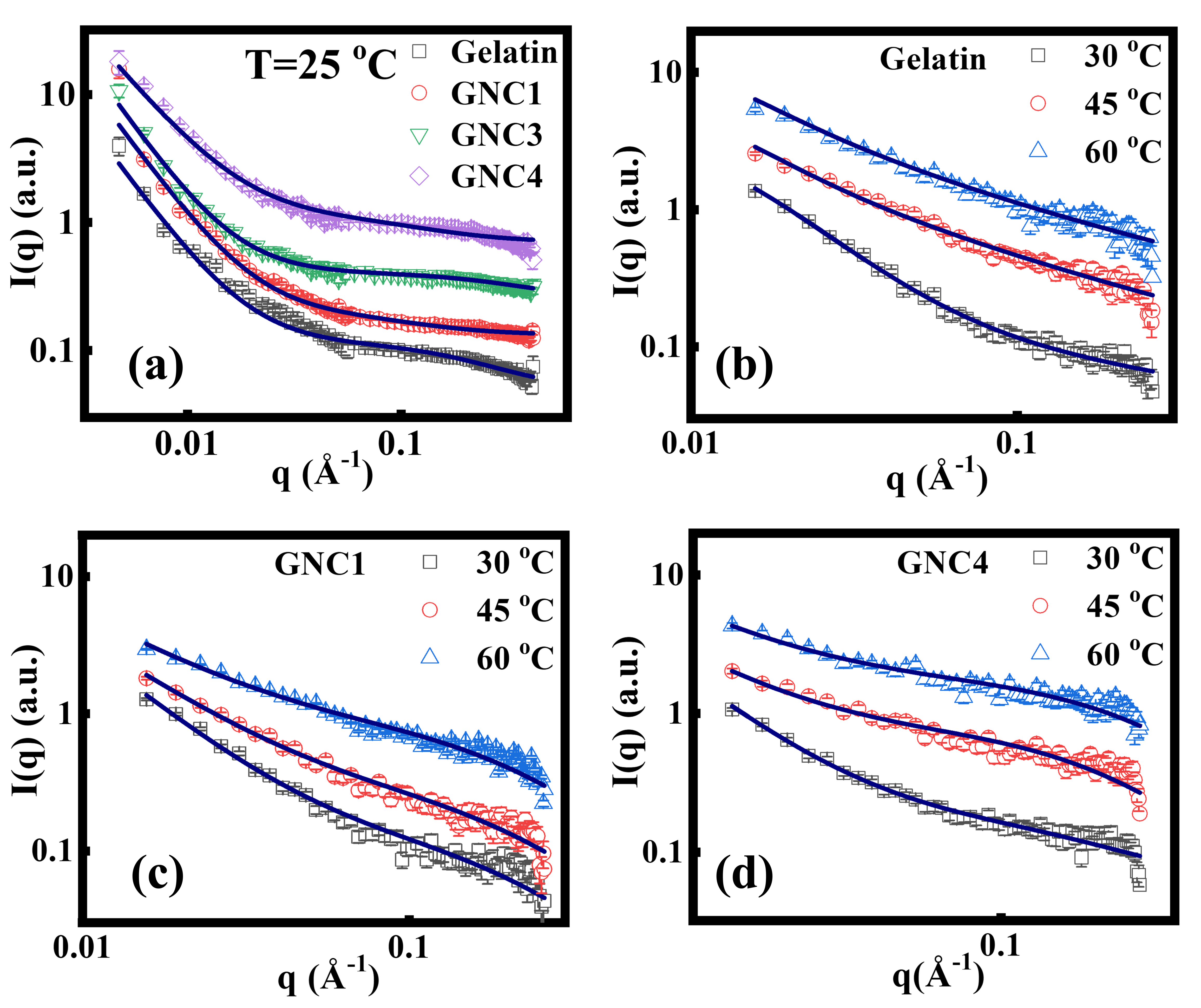}
\caption{SANS intensity profiles for gelatin hydrogel and composite hydrogels containing varying concentrations of \textit{PAuNRs} at different temperatures. (a) Room temperature data exhibits a dominant low-q slope for Gelatin, which decreases with increasing \textit{PAuNRs} concentration (GNC1 > GNC3 > GNC4), suggesting structural changes induced by \textit{PAuNRs} incorporation. (b, c, d) Temperature-dependent measurements for Gelatin, GNC1, and GNC4 reveal a consistent decrease in the low-q slope with rising temperature, indicating temperature-assisted structural modulation in the gelatin and gelatin composite hydrogels. The symbols used in plots (a), (b), and (c) are identical.}
  
  \label{Fig2}
\end{figure}

In Fig. 2(a), we show the scattering profile from the composite,  at different composition of \textit{PAuNR}. The data at varying temperature for neat hydrogel and composite of hydrogel and \textit{PAuNR} is presented in Fig. 2(b-d). The strong scattering at low wave vector in the \textit{I(q)} vs \textit{q} profile,  is evident for all samples, suggesting the presence of large-microscale structures within the hydrogels. The measured change in the slope of the scattering profile correlate with variations in \textit{PAuNR} composition and temperature, suggesting reorganisation in the hydrogel's internal network structure. The \textit{SANS} data for the hydrogel and composite samples, can be explained by combination of scattering  from two regions: long-range frozen inhomogeneities (cross-link rich region) and local liquid-like region (cross-link poor region)\cite{shibayama2012structure}. A two-stage model was employed to quantify the structural change in hydrogel-composite samples with \textit{PAuNR} composition and temperature variation\cite{bode2013hybrid}

\begin{equation}\label{first_eq}
I(q)=\frac{A}{q^{d_f}}+\frac{B}{{1+{\xi}^2q^2}}
\end{equation}\\

here \textit{A} and \textit{B} are scaling constants, $d_f$ corresponds to the fractal dimension of long-range inhomogeneities, $\xi$ represents the correlation length of density fluctuations and is related to the mesh size of the entangled networks. Theoretically, the estimate of fractal dimension, $d_f$, of \textit{1} corresponds to linear objects, \textit{1.67} corresponds to swollen polymer coils in good solvents, and \textit{2~-~3} corresponds to branched systems, such as chemical gels \cite{wei2021characterizing, adam1993fractal, boyd2021shining}.  \textit{Table. 1} reveals the fractal dimensions and correlation lengths obtained by fitting the gelatin hydrogel scattering data in Fig. 2 to Equation. 1. At 25 $^\circ\mathrm{C}$, the pristine gelatin hydrogel exhibits $d_f$ of $\approx 2.3$ indicating an inhomogeneity within the hydrogel structure. Interestingly, incorporating \textit{PAuNR} nanoparticles into the hydrogel (GNC1, GNC3, and GNC4) results in  decrease in the fractal dimension, with estimates of \textit{2.2}, \textit{2.1}, and \textit{2.0} respectively. The observed decrease in fractal dimension ($d_f$) upon incorporation of \textit{PAuNRs} occurs due to reduced interconnection between polymer network structures \cite{boyd2021shining,gubaidullin2022modulation}. This observation suggests that \textit{PAuNRs} disrupt the long-range inhomogeneities within the hydrogel networks, and promotes formation of a more uniform structure \cite{gubaidullin2022modulation}. The correlation lengths, estimated by fitting the data in Fig. 2 to Equation. 1, were found to be \textit{0.8 nm}, \textit{0.9 nm}, \textit{1 nm}, and \textit{1.1 nm} for pure gelatin, GNC1, GNC3, and GNC4, respectively. The correlation length represents the average distance over which structural features are correlated; thus a larger value suggests a more extended and less densely packed network. Thus the progressive increase in correlation length with increasing \textit{PAuNR} content indicates that \textit{PAuNR} incorporation reduces the compactness of the hydrogel network.\par
Further, we examined how temperature affects the arrangement of structures at the nanoscale in both the composite and pure gelatin samples. The $d_f$ estimates of pristine gelatin hydrogel decreases from \textit{2.0} at 30 $^\circ\mathrm{C}$ to \textit{1.75} at 45 $^\circ\mathrm{C}$ and \textit{1.5} at 60 $^\circ\mathrm{C}$ [\textit{Table S1}]. This suggests that the inhomogeneity within the hydrogel structure gets smaller as the temperature increases. Similar trends were observed for the GNC1 and GNC4 composite hydrogels.

\begin{table}[H]
    \centering
    \begin{tabular}{|c|c|c|} \hline 
 \multicolumn{3}{|c|}{\textbf{Temperature 25 $^\circ\mathrm{C}$}}\\ \hline 
         \textbf{Samples}&  \textbf{Fractal dimension}& \textbf{Correlation length (nm)}\\ \hline 
         Gelatin&  2.3& 0.8 \\ \hline 
         GNC1&  2.2& 0.9 \\ \hline 
         GNC3&  2.1& 1.0\\ \hline 
         GNC4&  2.0& 1.1\\ \hline
    \end{tabular}
    \caption{Fractal dimension and correlation length, at various \textit{PAuNR} composition obtained from two-stage model fitting to \textit{SANS} data in Fig.2.  \cite{valencia2020multivalent,bode2013hybrid}}
    \label{Table 1}
\end{table}

\subsection{Rheo-mechanical properties of hydrogel-\textit{PAuNR} composites}
To explore the relationship between the nanoscale structural changes induced by \textit{PAuNR} composition and the mechanical properties of the hydrogel composite, we performed rheological measurements using a  rheometer (\textit{HR20}, TA instruments) with a cone-plate geometry (diameter 20 mm, gap 26 micrometers). The rheological measurements include a strain sweep test ($0.01\%$ to $1000\%$ at frequency of 10 rad/sec) to assess the material's response to increasing deformation,  a frequency sweep test (0.1 to 100 rad$s^{-1}$ at strain of 10\%) to probe its viscoelastic behavior, a temperature sweep test (25 $^\circ\mathrm{C}$ to 60 $^\circ\mathrm{C}$, at 2 $^\circ\mathrm{C}$/minute, 10\% strain and 1 rad/sec) to investigate property changes with temperature, and a steady shear test ($0.1~s^{-1}$ to $500~s^{-1}$ shear rate) to understand its flow behavior under stress. Solvent trap accessories were employed in all tests to minimize solvent evaporation.\par

\begin{figure}[h!]
\centering
\includegraphics[width={\textwidth}]{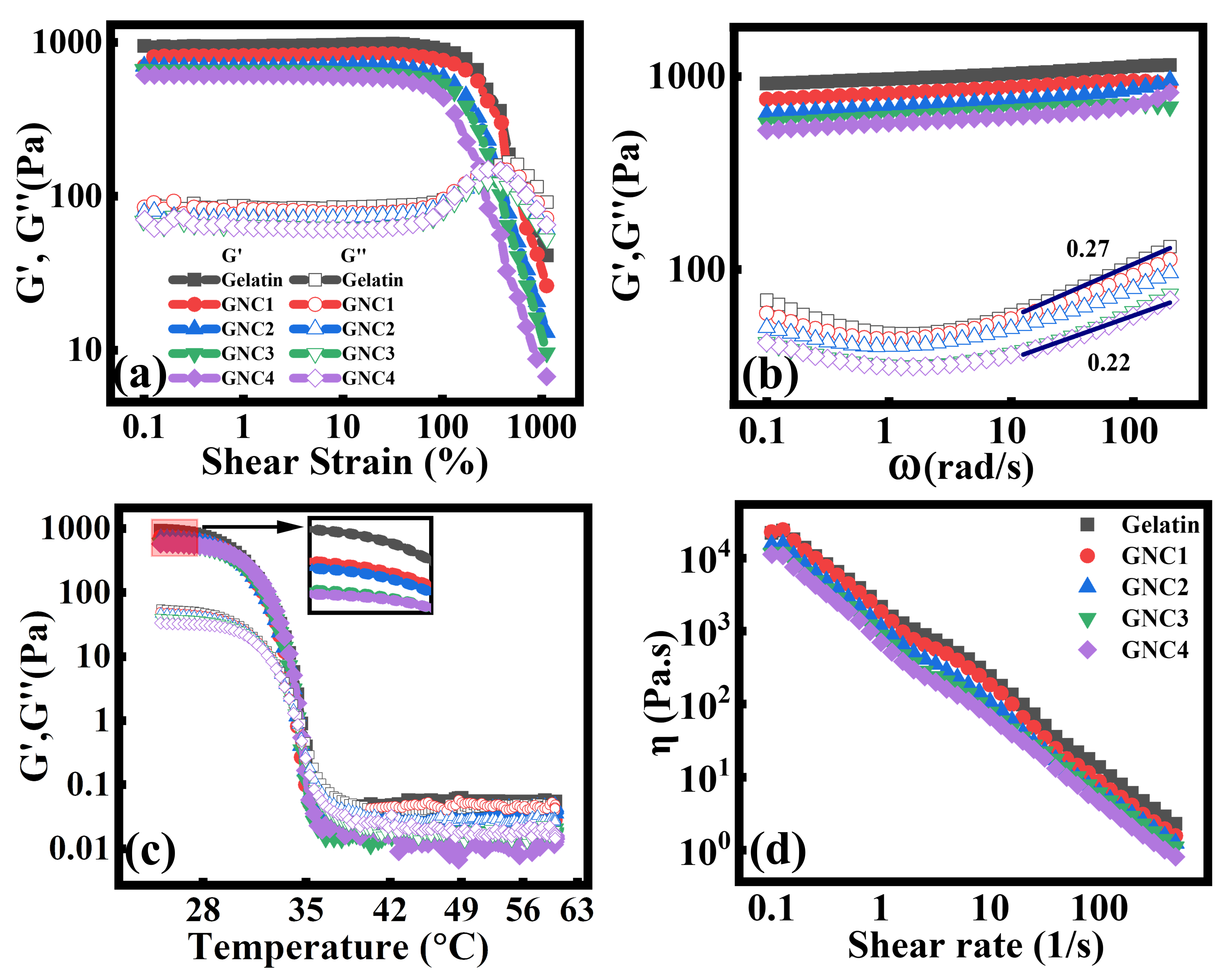}
\caption{Rheological measurements for gelatin and composite hydrogels: (a) The amplitude sweep reveals a strain-independent region where both G$^{'}$ and G$^{''}$ remain constant. With further increases in strain,  G$^{'}$ and G$^{''}$ decrease, eventually leading to structural breakdown. (b) The frequency sweep demonstrates the dominance of G$^{'}$ over  G$^{''}$ at all frequencies, indicating a solid-like gel behavior. The slope for $G^{''}~vs~\omega$ is indicated along side the data. (c) Temperature sweep data shows a decrease in both  G$^{'}$ and G$^{''}$ with increasing temperature, eventually resulting in a gel-to-sol transition at approximately 34.5 $^\circ\mathrm{C}$. (d) The shear flow test showed a decrease in $\eta$ with increasing shear rate, indicating shear-thinning behavior in all hydrogels. At the intermediate shear rate ($\approx ~ 1s^{-1}$ to  $20s^{-1}$), a shoulder in the viscosity curve suggested the presence of a local shear-thickening region.}
  \label{Fig3}
\end{figure}

The strain sweep data showing elastic,G$^{'}$, and loss,G$^{''}$, shear modulus at varying strains are shown in Fig. 3(a). The elastic shear modulus, G$^{'}$ represents the elastic energy stored within the gel structure, while the loss shear modulus, G$^{''}$ corresponds to the energy dissipated as the gel deforms \cite{SolichPRL20, tian2020construction,SSJACS2014,SrivastavaSM2011}. It is evident from the data that at low shear strain, G$^{'}$ and G$^{''}$ are independent applied strain . Beyond the critical yield strain,$\gamma_o$, (signifying irreversible material deformation), G$^{'}$ decreases with increasing shear strain. However the viscous modulus, G$^{''}$, reaches a peak value followed by decrease with increasing strain. The peak in G$^{''}$ reflects the breakdown of polymeric micro structures, a characteristic behavior observed in gel-like soft materials\cite{donley2020elucidating,tian2020construction}. At large shear strain (> $\gamma_o$) both G$^{'}$ and G$^{''}$ decreases in magnitude due to  alignment of polymers along the shear direction. \\
In Fig. 3(b), we show the results of a frequency sweep experiment, where G$^{'}$ and G$^{''}$ are plotted at varying angular frequency, $\omega$, for different composition of nanoparticle in the composite hydrogel. For all samples we measure  G$^{'}~\gg G^{''}$ (Fig. 3b), indicating that the hydrogels behave more like elastic solids than viscous liquids \cite{tian2020construction,SrivastavaSM2011}.  Notably, \textit{G$^{'}$} remains nearly independent of frequency (G$^{'}\propto\omega^0$), suggesting the formation of strong  networks within the gel\cite{SolichPRL20, SSLangmuir2022}, however, G$^{''}$ exhibits non-monotonous dependence on frequency.  We find that G$^{''}$ initially decreases, followed by a crossover point at intermediate frequency range, beyond which the G$^{''}$ vs $\omega$ curve shows increase in magnitude. In the large frequency regime, we measure G$^{''}~\propto~\omega^{0.24 }$ as shown in Fig. 3b.  In general, the power law exponent, \textit{n}, defined for G$^{''}~vs~\omega^{n}$ graph, is an characteristic feature to distinguish the material property\cite{SrivastavaSM2011} Typically for viscoelastic solids,  \textit{n = 1}, and  G$^{'}$$\propto\omega^0$ for all frequency ranges \cite{SSLangmuir2022} . For viscoelastic fluids, for low frequency regime, \textit{n = 1}, and G$^{'}$$\propto\omega^2$, with G$^{'}~\leq~G^{''}$; and at high frequency regime, \textit{n = 1}, with G$^{'}\propto\omega^0$ and G$^{'}~\geq~G^{''}$. For gelatin-\textit{PAuNR} composite gels, \textit{n = 0.24} ; $G^{'}$$\propto\omega^0$ and  G$^{'}~\geq~G^{''}$,  for all measured frequency range, signifying a elastic-gel like behaviour\cite{ roberto2023introduction,SolichPRL20}. The increase in \textit{G$^{''}$} at higher frequencies is attributed to the short-range movements of the polymer chains within the gel network \cite{du2004nanotube}. It can also be noted that the incorporation of \textit{PAuNRs} into gelatin hydrogels weakens their elastic properties as measured by the decrease in the estimates of \textit{G$^{'}$} for composite samples [Fig. 3(a-c)] . Pure gelatin exhibits highest \textit{G$^{'}$} value of \textit{1032 Pa}, whereas the \textit{G$^{'}$} for the GNC4 composite hydrogel is measured to be \textit{620 Pa} (Fig. S2a).  This decrease in \textit{G’} for the composite hydrogels is linked with lower fractal dimension networks\cite{gubaidullin2022modulation} also confirmed by \textit{SANS} analysis (Table.1).\par

Gelatin gels form a strong, three-dimensional structure at low temperatures due to interconnected polymer chains \cite{avallone2021gelation} and hydrogel exhibits predominant elastic behavior ($G^{'}~\gg~G^{''}$). As temperature increases, these chains loosen, resulting in disruption of the ordered arrangement of gelatin polymers inside hydrogel matrix\cite{zhang2017self} and unwind into a random coil like structure, causing the gel to transition into a liquid-like state, resulting in gel-to-sol transition. At $T\gg ~T_{gel-sol}$ (gel-to-sol transition temperature), temperature-assisted mobility of random coils of gelatin polymers leads to hydrogel in a viscous liquid state  and exhibits (G$^{'}$~<~G$^{''}$) \cite{zhang2017self}. The thermoresponsive behavior of gelatin makes it an promising candidate for bioprinting applications where temperature-controlled material properties are desired \cite{song2020injectable}.The gel-to-sol transition temperature, $T_{gel-sol}$, for our samples was investigated using a temperature sweep test (Fig. 3c). The critical $T_{gel-sol}$, was found to be around 34.5 $^\circ\mathrm{C}$ ± 0.5 $^\circ\mathrm{C}$ for the pure gelatin hydrogel. Interestingly, we don't observe any change in $T_{gel-sol}$  for composite samples and it was found to be independent of \textit{PAuNR} composition (Fig. S2b). The  $T_{gel-sol}$ of a hydrogel is influenced by the concentration of gelatin polymers \cite{osorio2007effects}  and  the presence of new interactions between gelatin and the nanocomponents used as fillers \cite{liu2020tunable,chen2022strong}. The nearly same $T_{gel-sol}$ for gelatin and composite hydrogel at fixed gelatin concentration (5\%) indicates the physical entanglement of gelatin polymers is solely responsible for the thermal stability of the hydrogel and is unaafected by the nanorod composition. \par
Flow behavior of gelatin hydrogel are characterized by analyzing the viscosity profile in the shear rate range of 0.1 $s^{-1}$ to 500 $s^{-1}$. Shear thinning behavior is evident for pure and composite hydrogel as viscosity decreases with increase in shear rate (Fig. 3d). The observation of shear thinning behavior suggests that the rate of disentanglement of the  polymer network due to applied shear stress is higher than the reconstruction rate of the network \cite{jiang2020rheological}. Interestingly, in the intermediate shear rate ($\approx ~ 1s^{-1}$ to  $20s^{-1}$), we measure an deviation from the shear-thinning behavior. This suggests a possible interplay between two opposing effects: shear-induced thinning and localized shear thickening. The localized thickening behavior might be attributed to the application of shear rate disrupting and opening polymeric clusters within the hydrogel. This disruption exposes hidden bonding sites, facilitating the formation of new networks between polymer chains, and leading to a temporary thickening effect\cite{wee2015cation}. In Fig. S3, local shear thickening behavior is shown for the observed range of shear rate. The local shear thickening effect is found to diminish with increasing composition of \textit{PAuNR} suggesting the presence of \textit{PAuNR} hinders the reconstruction of new network formation between polymers.\\
In Fig. 4a, we show the estimates of  critical yield strain,$\gamma_o$  and  viscosity, $\eta$, for pure gelatin and composite samples at various added  \textit{PAuNR} composition. It is evident that the composite samples exhibits lower values suggesting sample softening due to presence of  \textit{PAuNR}.  These observation suggests a potential disruption in network formation between gelatin polymers caused by the incorporation of \textit{PAuNRs} within the matrix. To further analyze the mechanical properties of the gelatin and gelatin composite hydrogels, uniaxial compression tests were conducted using a rheometer (Anton Paar MCR 702) on cylindrical hydrogel samples with an 8 mm diameter and 4 mm height, using an 8 mm parallel plate geometry [\textit{SI}]. The compression strength and fracture strain of the gelatin and gelatin composite hydrogels, as presented in Fig. S2(d), were determined from the stress-strain curve shown in  (Fig. S2(c). The pure gelatin hydrogel exhibited a compression strength of 0.0243 MPa and a fracture strain of 80.53\%. In contrast, the incorporation of \textit{PAuNR} led to a decrease in both compression strength and fracture strain, with GNC4 showing the lowest values of 0.0107 MPa and 67.20\%, respectively. This reduction is attributed to the decreased entanglement between gelatin polymers due to the presence of \textit{PAuNR} \cite{kim2022reactive}. These findings are consistent with the results from shear rheology tests, indicating a decrease in mechanical stability with \textit{PAuNR} incorporation.  As for future application, the shear thinning behavior of this gelatin composite hydrogel system makes it suitable for extrusion-based 3D-bioprinting as well in easier surgical placement \cite{gockler2021tuning}.\\ 

\subsection{Differential scanning calorimetry (DSC) analysis}
The denaturation temperture, $T_d$, of the gelatin and gelatin composite hydrogels was investigated using  differential scanning calorimeter (Hitachi model 7020) to validate the concept of the reduced entanglement in the polymer gelatin network. Gelatin is the denatured form of collagen, with its protein chains existing primarily in random coil configurations. However, during the gelatin processing, some polypeptide intermolecular hydrogen bonds between carbonyl oxygen and amide hydrogen can reform and causing them to renature back into the triple helix structure \cite{zhang2006crosslinking,wu2018ionic,SSPRL2007}. The endothermic peak observed in Fig. S4 corresponds to the denaturation process of gelatin's triple helix polymers, where they unfold into random coils. The temperature at endothermic peak represents the denaturation temperature $T_{d}$ \cite{mosleh2021structure}. For measurement,  $\approx~4-5mg$ of ample was added into an aluminum hermetic pan and heated from 30 $^\circ\mathrm{C}$ to 300 $^\circ\mathrm{C}$ at a rate of 10 $^\circ\mathrm{C}$/minute in a nitrogen environment. \par
The $T_d$ estimate at varying composition of \textit{PAuNR} for different composite samples is shown in Fig. 4b. The DSC thermogram of gelatin and gelatin nanocomposite hydrogel exhibits an distinct endothermic peaks as shown in Fig. S4.  The $T_d$ estimate for pure gelatin hydrogel is 134 $^\circ\mathrm{C}$ which is found to progressively decrease with increasing \textit{PAuNR} concentration, reaching its lowest value of 103 $^\circ\mathrm{C}$ for GNC4 (Fig. 4b). The observed decrease in $T_d$, suggests that \textit{PAuNR} incorporation hinders the renaturation of triple helices within the gelatin hydrogel. As due to the presence of \textit{PAuNRs} within the gelatin matrix, hindered the formation of cross-links between gelatin polymer chains. This decrease in cross-linking density weakens the overall network structure of the hydrogel. Consequently, the hydrogel exhibits lower resistance to the temperature-induced unfolding (denaturation) of the triple-helix structure of the gelatin polymer \cite{yu2020preparation}. Incorporation of \textit{PAuNRs} inside gelatin hydrogel matrix hampers networks formation resulting in lower thermal stability and weak gel strength \cite{youssefi2021green,SSPRL2007}. 

\begin{figure}[h!]
\centering
\includegraphics[width={\textwidth}]{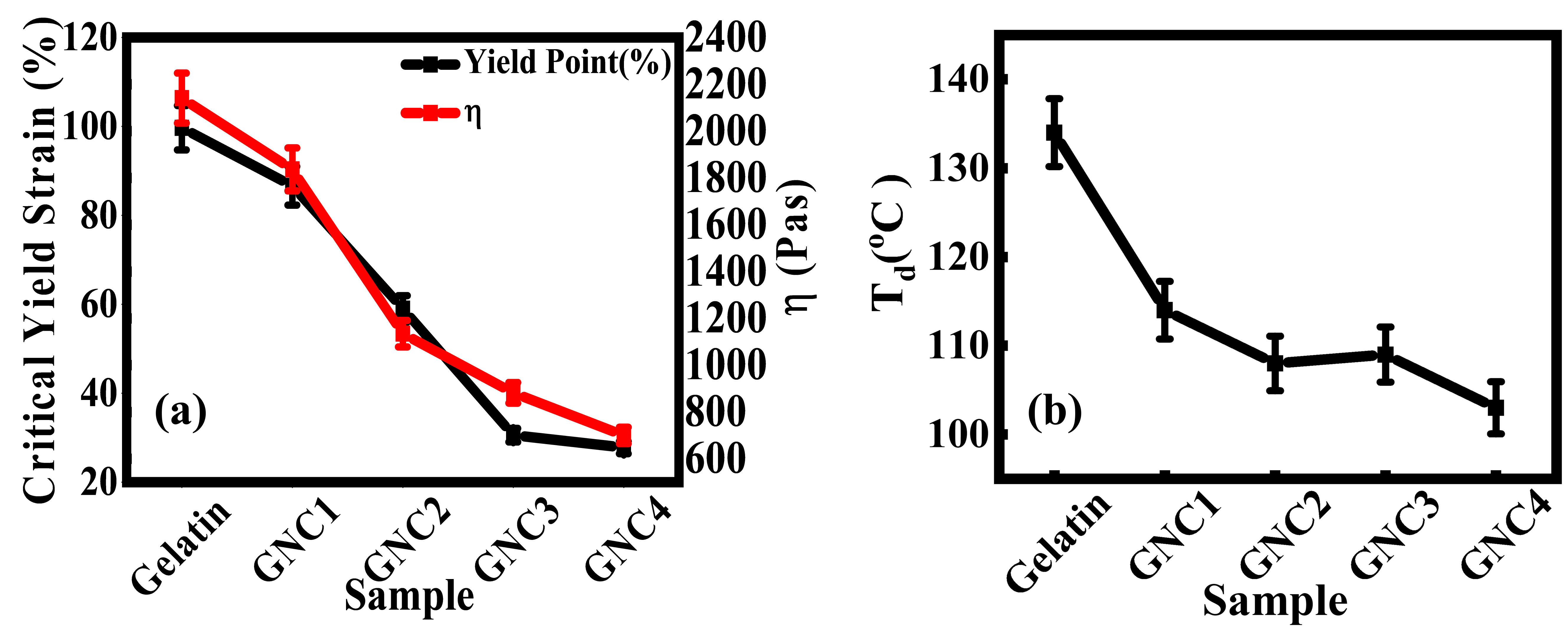}
\caption{Effect of \textit{PAuNR} concentration on hydrogel properties: (a) Critical yield strain, $\gamma_o$ and viscosity, $\eta$ (at shear rate of $1s^{-1}$) were found to decrease with increasing \textit{PAuNR} concentration in the composite hydrogel. (b) Denaturation temperature ($T_d$) obtained from DSC analysis, indicating a decrease in $T_d$ with the incorporation of \textit{PAuNR}. }
\label{Fig4}
\end{figure}
\subsection{Swelling Properties}
To test the stability of the hydrogel inside the aqueous environment, we conducted swelling experiment. Hydrogel samples molded in cylindrical shape of diameter $8~mm$ and height $4~mm$ were immersed in Millipore water at 25 $^\circ\mathrm{C}$ for 72 hours and their weight was monitored at regular intervals. The swelling ratio of hydrogel samples was studied by the gravimetric method. At specific time intervals of 0.5, 1, 1.5, 2, 10, 24, 48, and 72 hr, the swollen hydrogel was weighed after removal of surface water using filter paper. The swelling ratio (SR), is calculated using the formula,  SR=$\frac{w_t - w_o}{w_o}*100$. Here, ${w_o}$, refers as the initial weight of hydrogel before immersion and ${w_t}$ represents the swollen hydrogel weight after time $t$. Here $w_o$ is the initial weight of the hydrogel and $w_t$ represents the real-time weight during solvent absorption.

\begin{figure}[h!]
\centering
\includegraphics[width={\textwidth}]{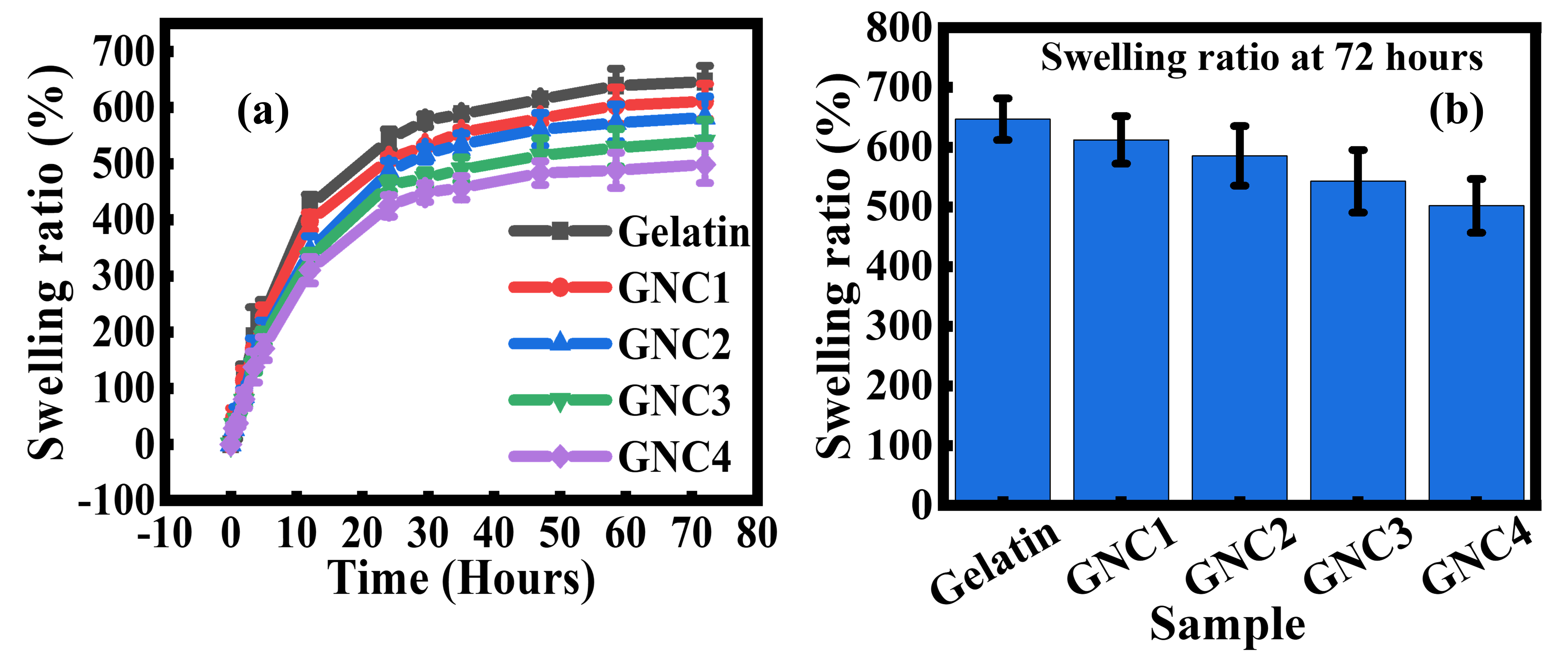}
  \caption{Swelling Properties of Gelatin and Composite Hydrogels: (a) Swelling ratio of both hydrogels, showing a rapid increase up to 10 hours, followed by a gradual rise until reaching a plateau at approximately 36 hours, with the highest swelling ratio observed at 72 hours. (b) The swelling ratio at 72 hours, indicating a decrease in the swelling ratio of hydrogel with increasing \textit{PAuNR} concentration.}
  \label{Fig5}
\end{figure}

Fig. 5(a) illustrates the swelling behavior of all hydrogels over time. The swelling ratio (SR) of each hydrogel initially exhibited a rapid increase within the first 10 hours, followed by a gradual rise until reaching a plateau at approximately 36 hours, suggesting that the hydrogels had reached their maximum swelling capacity. While pure gelatin hydrogels exhibit the highest swelling ratio (644$\%$) at 72 hours (Fig. 5(b)), incorporating \textit{PAuNRs} progressively reduces water uptake (down to 499$\%$ for GNC4). This decrease is likely due to \textit{PAuNRs} hindering water diffusion  and reducing the number of hydrophilic groups available for interaction thus reducing solvent diffusion through capillary action\cite{daniel2013effects,li2003gelatin}, ultimately leading to a more stable hydrogel in aqueous environments.

\section{CONCLUSION}
We present the experimental studies on structure-property correlation in thermoresponsive gelatin hydrogels containing varying amounts of light-sensitive antimicrobial \textit{PAuNRs}, at various temperatures. Our analysis revealed that \textit{PAuNR} addition transforms the hydrogel's surface structure from smooth to porous, and results in reduction in internal structural complexity, as evident from decrease in fractal dimension and an increase in correlation length. Further, the effect of structural modulation on the bulk properties of gelatin composite hydrogel is investigated using the changes in storage moduli, viscosity, critical yield strain,  denaturation temperature and swelling ratio of the hydrogels. The effect of temperature on the structure and property of hydrogel is also analyzed by comparing its fractal dimension and storage modulus. An increase in temperature decreases the fractal dimension as well as the storage moduli for both gelatin and gelatin composite hydrogel. This revealed that change in the stability of hydrogel depends on \textit{PAuNR} and temperature-assisted structure modulation. The reduced elasticity of the composite hydrogel and light sensitive/antimicrobial property of the  \textit{PAuNRs} makes this system desirable for specific biomedical applications, such as tissue engineering, device fabrication and stimuli based controlled drug delivery devices respectively. This work unravels the effect of \textit{PAuNR} and temperature on the structure and properties of gelatin composite hydrogels and paves the way for the development of advanced materials with tailored properties for specific applications.  

\section*{SUPPLEMENTARY MATERIAL}
The supplementary material provides additional data for characterization of gelatin and gelatin composite hydrogels. It includes a table with details of the fractal dimensions at various temperatures. Additionally,  data on the storage modulus and gel-to-sol transition temperature are provided. The supplementary material also features results from compression tests assessing the mechanical strength, plots demonstrating local shear thickening behavior. Differential Scanning Calorimetry (DSC) thermograms illustrating the thermal properties of the hydrogels are shown.

\begin{acknowledgments}
IK acknowledges financial support from the UGC, India. SS acknowledges support from the CRS-UGC-DAE, India. We thank the SAIF-IITB, the Central Facility of Chemical Engineering-IITB, and the Monash Instrumentation Facilities-IITB for their assistance with material characterization. We also thank the Bhabha Atomic Research Centre (BARC), Mumbai, for access to Neutron Scattering facilities.

\end{acknowledgments}

\section*{AUTHOR DECLARATIONS}
\subsection*{Conflict of Interest}
The authors have no conflicts to disclose.

\section*{Data Availability Statement}
The data that support the findings of this study are available from the corresponding author upon reasonable request.

\bibliography{references}

\end{document}


\preprint{Preprint}

\title{Supporting Information: A Composite Hydrogel of Porous Gold Nanorods and Gelatin: Nanoscale Structure and Rheo-Mechanical Properties}

\author{Irfan Khan}
\affiliation{ 
Department of Physics, Indian Institute of Technology Bombay
}
\author{Snigdharani Panda}
\affiliation{ 
Department of Physics, Indian Institute of Technology Bombay
}
\author{Sugam Kumar}
\affiliation{ 
Solid State Physics Division, Bhabha Atomic Research Centre
}
\author{Sunita Srivastava*}
 \email[Author to whom correspondence should be addressed:]{sunita.srivastava@iitb.ac.in}
\affiliation{ 
Department of Physics, Indian Institute of Technology Bombay
}%

\date{\today}

\maketitle

\section{\label{sec:level1} Dimensional Analysis of \textit{PAuNRs}}
The dimension of \textit{PAuNR} was measured using ImageJ software, that works on concept of pixel-based calibration on SEM images of langth and width. The data obtained from imageJ software were further processed in Origin software for model fitting. A Gaussian fitting method was applied to the length and width distribution of over 300 nanorods. This approach provided the mean nanorod length of $140 \pm 11 \, \text{nm}$ and width of $67 \pm 6 \, \text{nm}$ (Fig. S1 (a) and (b)), indicating a narrow distribution of \textit{PAuNR} sizes.

\begin{figure}[H]
\centering
\includegraphics[width={\textwidth}]{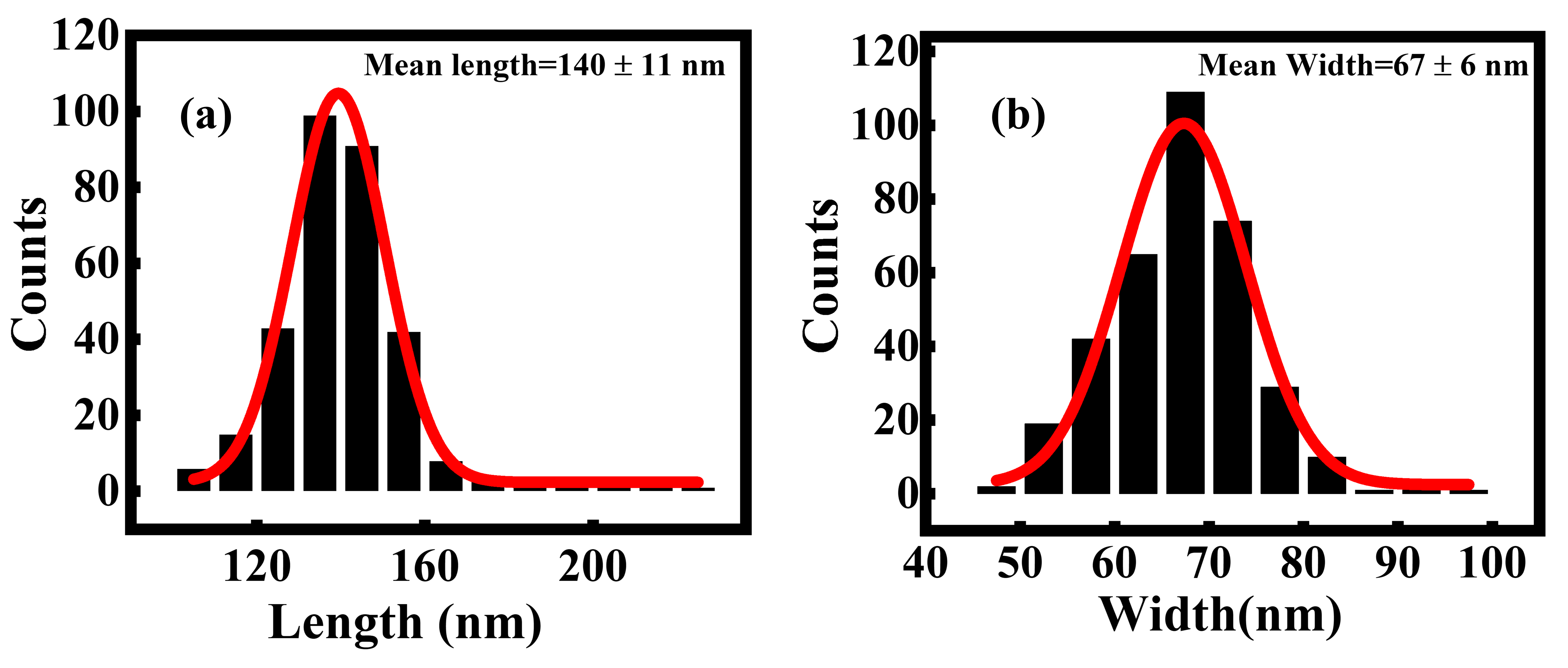}
\caption{ SEM image analysis for \textit{PAuNRs} using gaussian fitting (a) Plot of length distribution of \textit{PAuNR} with mean length of $140 \pm 11 \text{nm}$ (b) Plot of width distribution of \textit{PAuNR} with mean width of  $67 \pm 6 \text{nm}$.}
\label{fig.S1}
\end{figure}

\section{Estimates of fractal dimension at various temperature from \textit{SANS} data}
The \textit{I(q)} vs \textit{q}, profile were fitted using Equation. 1, as described in the main text. The fractal dimension, $d_f$ for composite hydrogel with varying \textit{PAuNRs} composition at various temperature is shown in the table below. 
\begin{table}[H]
    \centering
    \begin{tabular}{|c|c|c|c|} \hline 
         \textbf{Temperature}&  \textbf{Gelatin}& \textbf{GNC1}&\textbf{GNC4}\\ \hline 
         $(^\circ\mathrm{C})$&  Fractal dimension& Fractal dimension&Fractal dimension\\ \hline 
         30 &  2.00& 1.90& 1.80\\ \hline 
         45 &  1.75& 1.60& 1.56\\ \hline 
         60 &  1.50& 1.40& 1.45\\ \hline
    \end{tabular}
    \caption{ Fractal dimension of gelatin and gelatin composite hydrogels (GNC1, GNC4) at indicated temperatures, derived from fitting \textit{SANS} line profiles to Equation.1.}
    \label{TABLE. S1}
\end{table}
\section{Rheo-mechanical and thermal Properties of Gelatin and Gelatin Composite Hydrogels}
The storage modulus (G$^{'}$) obtained from frequency sweep tests (Fig. S2(a)), illustrates the strength of gelatin and gelatin composite hydrogels. The data reveal a notable softening effect in the composite hydrogels, characterized by a decrease in G$^{'}$ with increasing \textit{PAuNR} composition. This reduction in G$^{'}$ indicates that the incorporation of \textit{PAuNR} leads to a less rigid hydrogel network. Hydrogel thermal propeties under shear were studied using the temperature sweep experiment (main Fig. 3(c)). The gel-to-sol transition temperature for the hydrogels was observed to be at a constant 34.5 $^{\circ}\text{C}$ irrespective of composition (Fig. S2(b)). The stress-strain curves from compression tests are shown in Fig. S2(c), illustrating the deformation behavior and stress responses of the hydrogels. Initially, the curves exhibit linear behavior at low compression strains. As compression increases, the curves transition to non-linear deformation until the sample reaches its compressive strength and subsequently fractures.The compressive strength and fracture strain derived from the stress-strain curves (Fig. S2(c)) are summarized in Fig. S2(d). Pure gelatin hydrogel exhibits superior compressive strength and fracture strain compared to composite hydrogels. The incorporation of \textit{PAuNR} results in a marked reduction in both compressive strength and fracture strain, with the GNC4 composite hydrogel demonstrating the lowest values. This decrease is attributed to reduced polymer entanglement and compromised network integrity due to the presence of \textit{PAuNR}.
\begin{figure}[H]
\centering
\includegraphics[width=\textwidth]{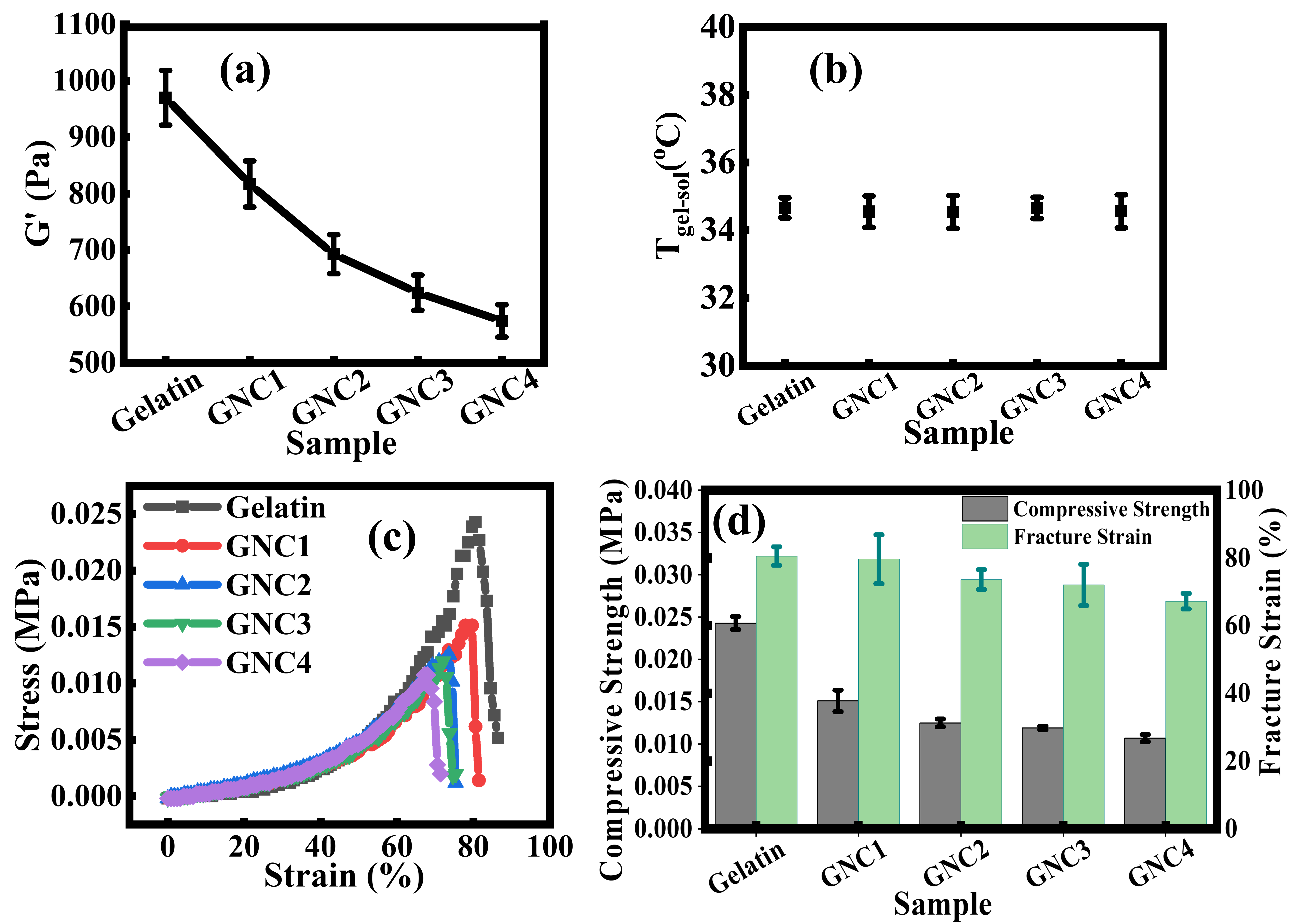}
  \caption{(a) Storage modulus (G$^{'}$) derived from frequency sweep test at 10 rad/s, is plotted for gelatin and gelatin composite hydrogel. The result indicates G$^{'}$ decreasing with PAuNR incorporation, as composite hydrogel shows lower gel strength. (b) Gel-to-sol transition temperature ($T_{gel-sol}$) obtained from temperature sweep test, plotted for gelatin and gelatin composite hydrogel.
  $T_{gel-sol}$ is observed to be at 34.5 $^{\circ}\text{C}$, regardless of \textit{PAuNRs} composition within the gel. (c) The stress versus strain curve is plotted for gelatin and gelatin composite samples. Stress rises initially linearly with strain, followed by non-linear, and finally to the yielding of the sample. (d) Compressive strength and fracture strain obtained from plot (c) are plotted for gelatin and gelatin composite hydrogel. It shows that both the parameters reduce with PAuNR incorporation, as composite hydrogel exhibits lower Compressive strength and fracture strain.}
  
  \label{Fig.S1}
\end{figure}
\section{Analysis of shear flow data }
The shear flow data to esimate viscosity, $\eta$, at varying shear rate is shown in Fig.3d [main text]. Broadly shear thinning behavior is observed as evident from decrease in viscosity with increasing shear rate. Interestingly, for small intermediate range of shear rate, shear thickening behavior has been observed as discussed in the main text. Baseline corrections are performed to highlight the shear thickening behavior shown by a well-defined peak in the data as shown in Fig. S2. It is evident from the data that the local shear thickening response is suppressed with increasing composition of \textit{PAuNRs}. For sample, GNC4, with highest composition of \textit{PAuNRs} marginal trace of shear thickening is present.
\begin{figure*}[h]
\centering
\includegraphics[width=0.5\textwidth]{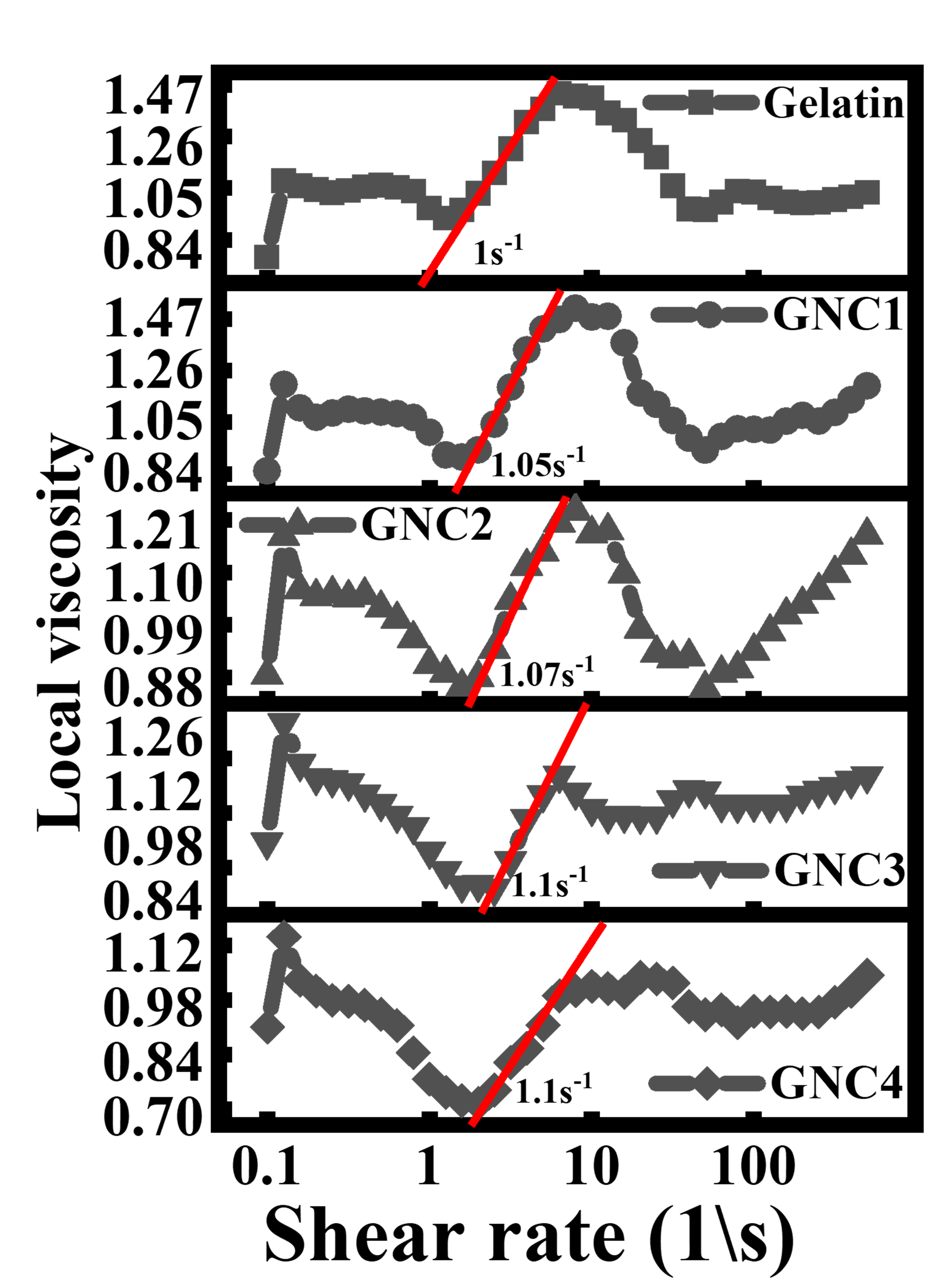}
\caption{Baseline correction analysis of $\eta~$ \textit{vs shear rate} data to highlight the shear thicknening response of pure and composite gelatin hydrogel, in the intermediate shear rate. The onset of this response is independent of the \textit{PAuNRs} composition. Local viscosity is dimensionless because it is obtained by dividing the viscosity by the linearly fitted viscosity.}
\label{Fig.S4}
\end{figure*}
\section{Differential Scanning Calorimetry (DSC) Analysis of Gelatin and Gelatin Composite Hydrogels} 
DSC was employed to investigate the thermal behavior of gelatin and gelatin composite hydrogels. The DSC measurements were conducted under an inert nitrogen atmosphere over a temperature range of 30 $^{\circ}\text{C}$ to 300 $^{\circ}\text{C}$ to ensure an accurate assessment of thermal transitions. The DSC thermograms for all samples, irrespective of composition, exhibit a prominent endothermic peak associated with the denaturation of gelatin polymers (Fig. S(4)). This peak signifies the temperature at which the gelatin structure undergoes thermal degradation. Notably, the position of the endothermic peak is highest for the pure gelatin hydrogel 134 $^{\circ}\text{C}$, reflecting its relatively greater thermal stability. As the concentration of \textit{PAuNR} increases within the hydrogel matrix, the endothermic peak shifts to progressively lower temperatures. The GNC4 composite hydrogel demonstrates the lowest peak position 103 $^{\circ}\text{C}$, indicating a significant reduction in thermal stability compared to pure gelatin. This observed trend suggests that \textit{PAuNR} incorporation affects the thermal stability of the hydrogel, due to interactions within the hydrogel matrix. The data underscores the need for careful consideration of \textit{PAuNR} concentration in applications where thermal stability is critical.
\begin{figure}[h]
\centering
\includegraphics[width=0.4\textwidth]{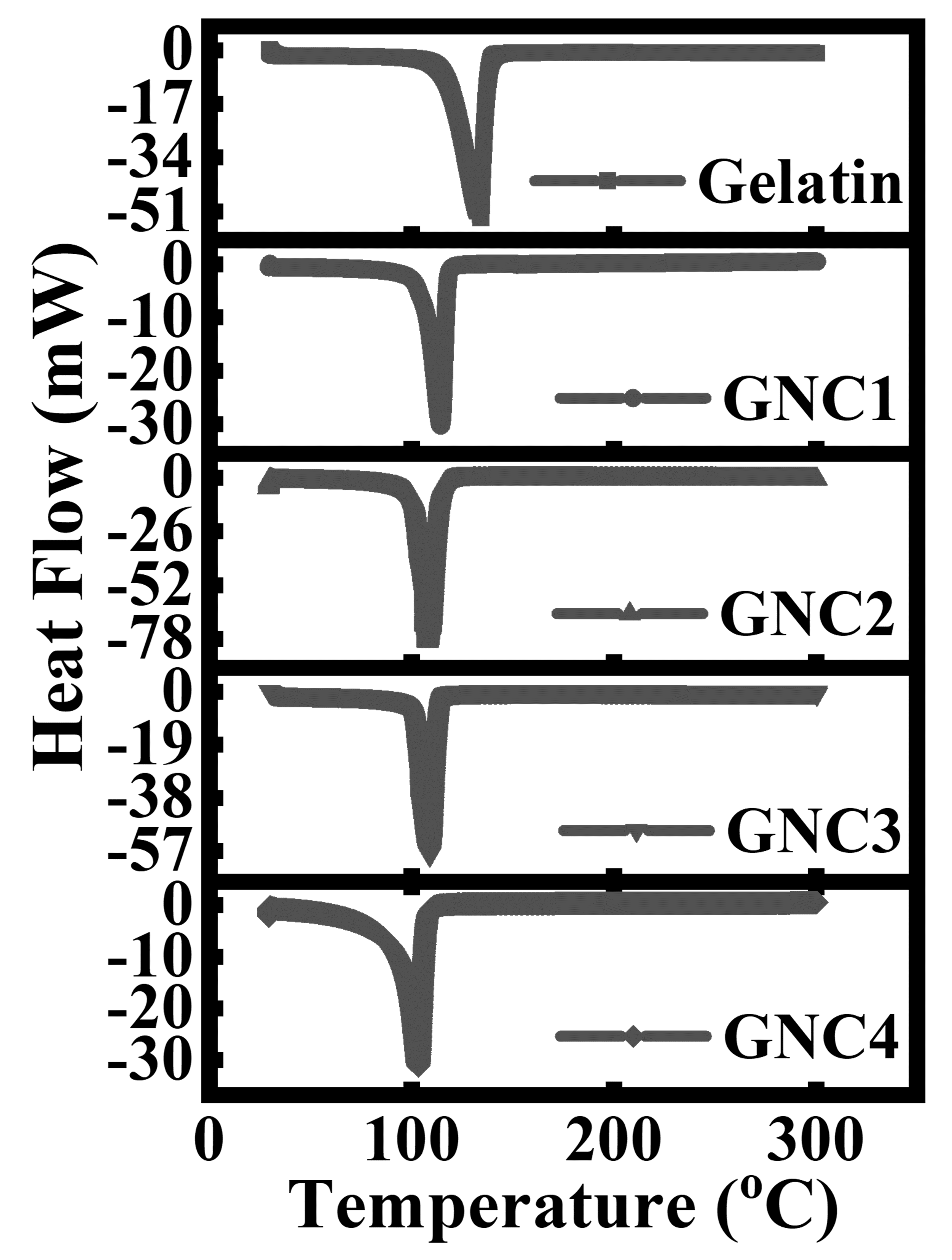} 
\caption{Differential Scanning Calorimetry (\textit{DSC}) thermogram reveal the thermal stability of gelatin and composite hydrogel. Endothermic peak is associated with denaturation of gelatin polymer with the corresponding denaturation temperatrue ($T_d$). Gelatin hydrogel exhibits the highest $T_d$, the presence of \textit{PAuNRs} lower $T_d$, with GNC4 showing lowest of all the $T_d$.}

\label{Fig.S5}
\end{figure}